\documentclass[aps,prl,twocolumn,superscriptaddress,showpacs]{revtex4-2}

\usepackage{graphicx}
\usepackage{amsmath,amssymb}
\usepackage{bm}
\usepackage{color}
\usepackage{indentfirst}

\makeatletter
\def\btt#1{\texttt{\@backslashchar#1}}%
\DeclareRobustCommand\bblash{\btt{\@backslashchar}}%
\makeatother

\begin{document}

%\baselineskip=2\normalbaselineskip
%↑１行空きにする場合

\title{Experimental Verification of Charge Soliton Excitations in the Ionic Mott-Peierls Ferroelectric, TTF-CA}

\author{R. Takehara}
\email{Corresponding author: takehara.r.ab@m.titech.ac.jp}
\affiliation{Department of Applied Physics, University of Tokyo, Bunkyo-ku, Tokyo 113-8656, Japan}

\author{H. Adachi}
\affiliation{Department of Applied Physics, University of Tokyo, Bunkyo-ku, Tokyo 113-8656, Japan}

\author{K. Sunami}
\affiliation{Department of Applied Physics, University of Tokyo, Bunkyo-ku, Tokyo 113-8656, Japan}

\author{K. Miyagawa}
\affiliation{Department of Applied Physics, University of Tokyo, Bunkyo-ku, Tokyo 113-8656, Japan}

\author{T. Miyamoto}
\affiliation{Department of Advanced Materials Science, University of Tokyo, Kashiwa, Chiba, 277-8561, Japan}

\author{H. Okamoto}
\affiliation{Department of Advanced Materials Science, University of Tokyo, Kashiwa, Chiba, 277-8561, Japan}
\affiliation{AIST-UTokyo Advanced Operando-Measurement Technology Open Innovation Laboratory (OPERANDO-OIL), National Institute of Advanced Industrial Science and Technology (AIST), Chiba 277-8568, Japan}

\author{S. Horiuchi}
\affiliation{Research Institute for Advanced Electronics and Photonics (RIAEP), National Institute of Advanced Industrial Science and Technology (AIST), Tsukuba, Ibaraki, 305-8565, Japan}

\author{K. Kanoda}
\email{Corresponding author: kanoda@ap.t.u-tokyo.ac.jp}
\affiliation{Department of Applied Physics, University of Tokyo, Bunkyo-ku, Tokyo 113-8656, Japan}

\date{\today}

\begin{abstract}
Strong coupling of charge, spin, and lattice in solids brings about emergent elementary excitations with their intertwining and, in one dimension, solitons are known as such.  The charge-transferred organic ferroelectric, TTF-CA, has been argued to host charge solitons; however, the existence of the charge solitons remains unverified. Here, we demonstrate that the charge-transport gap in the ionic Mott-Peierls insulating phase of TTF-CA is an order of magnitude smaller than expected from quasiparticle excitations, however, being entirely consistent with the charge soliton excitations. We further suggest that charge and spin solitons move with similar diffusion coefficients in accordance with their coexistence. These results provide a basis for the thermal excitations of the emergent solitons.
\end{abstract}


\maketitle

%\section{Introduction}
Topological excitations in one dimension (1D) are zero-dimensional defects behaving like particles. They are known as solitons and domain walls, which occasionally cause unconventional electrical and magnetic properties \cite{1, 2, 3, 4, 5, 6, 7, 8, 9, 10, 11, 12}. Notably, the solitons expected to emerge in the neutral-ionic (NI) transition material, tetrathiafulvalene-$\textit{p}$-chloranil (TTF-CA), are of profound interest in that they can be elementary excitations responsible for electrical and magnetic properties instead of electrons in a Mott-Peierls system, in which charge, spin, and lattice are strongly entangled \cite{13, 14, 15}. Furthermore, the solitons emergent at 1D ferroelectric boundary have been predicted to possess fractional charge \cite{16, 17, 18}.

In TTF-CA, an electron donor (D) molecule, TTF, and an acceptor (A) molecule, CA, alternately stack one-dimensionally [Fig. 1] \cite{19}. With applying pressure or lowering temperature, the neutral TTF-CA crystal progressively gains D-A electrostatic energy and then transitions or crosses over to an ionic Mott state by a charge transfer from TTF to CA  \cite{20, 21}. Ionicity is measured by the degree of charge transfer, $\rho$, which is represented by $\rho_{\mathrm{N}}$$\sim$0.25 in the neutral state and $\rho_{\mathrm{I}}$$\sim$0.75 in the ionic Mott state \cite{22, 23, 24, 25, 26, 27, 28}. Additionally, strong electron-lattice interactions cause static (dynamical) lattice dimerization along the 1D chain in a ferroelectric (paraelectric) ionic phase, denoted as $\mathrm{I_{ferro}}$ ($\mathrm{I_{para}}$) phase hereafter, by Peierls or spin-Peierls mechanism [Fig. 1] \cite{21, 29, 30, 31}. In the $\mathrm{I_{ferro}}$ phase, the dimerization of TTF and CA is 3D ferroelectric long-range ordered, whereas, in the $\mathrm{I_{para}}$ phase, thermal fluctuation of the dimerization breaks the long-range order, forming a “dimer liquid” state \cite{11, 31}.

In general, the resistivity of organic semiconductors substantially decreases and often becomes metallic with pressure increased to several kbar \cite{32}. However, in the $\mathrm{I_{para}}$ phase of TTF-CA, the resistivity at room temperature is insensitive to pressure in a vast pressure region from 20 kbar at least up to 80 kbar \cite{33}. Significantly, the $\mathrm{I_{para}}$ phase is suggested to host two types of solitons, charge and spin solitons, that emerge on the boundaries of the oppositely polarized domains [Fig. 1], both of which are predicted to contribute to electrical conduction \cite{12, 13, 14, 15}. The pressure-insensitive electrical conductivity possibly stems from the peculiar nature of the solitons; however, the experimental verification of the existence of the charge soliton has not yet been done. In the present work, to obtain experimental indications of the charge soliton excitations contributing to electrical conduction in TTF-CA under pressure, we have investigated the temperature and pressure dependence of electrical resistivity of TTF-CA. To illustrate the peculiar electrical transport nature of TTF-CA under pressure, we have also investigated, as a reference system, the similar ionic material, tetrathiafulvalene-$\textit{p}$-bromanil (TTF-BA), a nearly fully charge-transferred Mott insulator with $\rho_{\mathrm{I}}$ of $\sim$0.95 \cite{34, 35, 36}, in which the charge degrees of freedom is strongly suppressed. The comparison of the two systems brings to light the existence of thermally excited solitons and their contribution to electrical conduction. 

%%%%%%%%%%%%%%%%%%%Fig1
\begin{figure*}[htbp]
\centering
\includegraphics[width=17.5cm,clip]{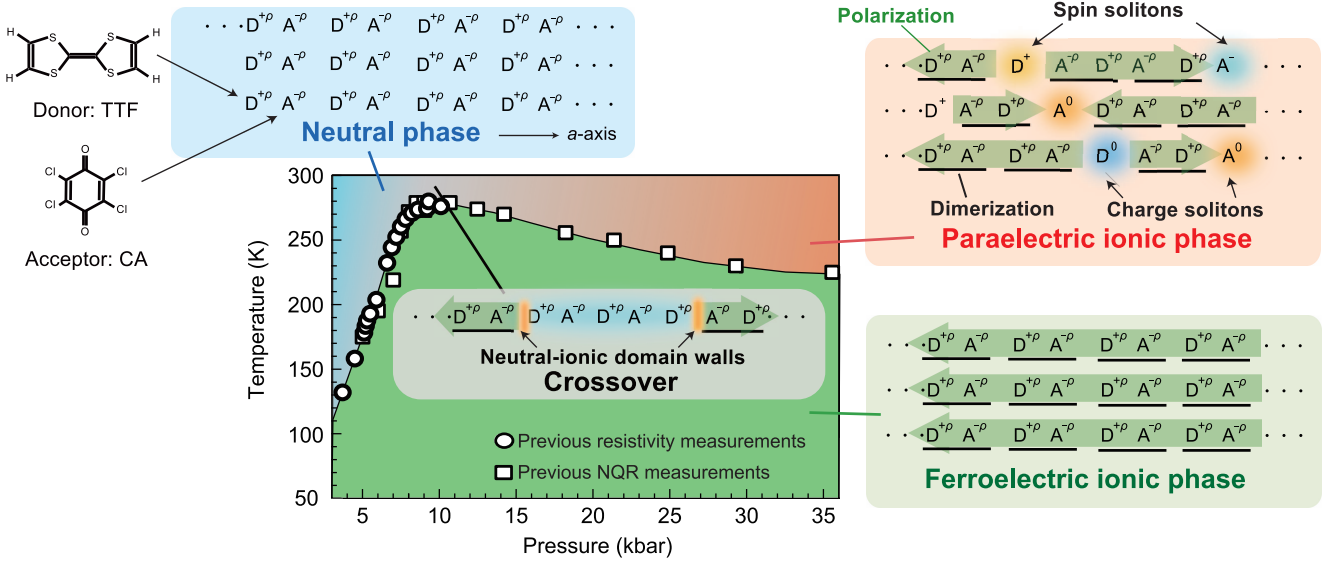}
\caption{Phase diagram of TTF-CA and schematic illustrations of the N, $\mathrm{I_{para}}$, and $\mathrm{I_{ferro}}$ phases with the charge transfer indicated by $\rho$ and the polarization of dimers pointed by bold green arrows. Mobile charge and spin solitons appear in the $\mathrm{I_{para}}$ phase. In the crossover between the N and $\mathrm{I_{para}}$ phases, NI domain walls are excited.
}
\label{Fig1} 
\end{figure*}
%%%%%%%%%%%%%%%%%%%Fig1

%\section{Experimental}
We performed the electrical resistivity measurements for TTF-CA and TTF-BA under pressure with the four-terminal method. The electrical current was injected along the 1D chains, namely, the $\textit{a}$-axis for TTF-CA and the $\textit{b}$-axis for TTF-BA. The samples were mounted in a clamp-type piston-cylinder pressure cell, and Daphne 7373 and 7474 oils were used as the pressure-transmitting media up to 20 kbar and 35 kbar, respectively, where pressure is hydrostatic \cite{37}. The pressure values quoted in this paper indicate internal pressures that were reduced from external pressures by the pressure-efficiency factor of 0.9 determined separately with a Manganin wire used as an indicator of the internal pressure.

%\section{Results and Discussion}
%\subsection{Pressure dependence of resistivity of TTF-CA and TTF-BA}
Figure 2 (a) compares the pressure dependences of the resistivities of TTF-CA and TTF-BA at room temperature. With increasing pressure, the resistivity of TTF-CA takes a minimum at 8-9 kbar and levels off to a value of $\sim$2.5 $\mathrm{\Omega}$cm above 20 kbar. Our previous study demonstrated that the resistivity minimum results from the NI domain wall (NIDW) excitations arising around NI crossover pressure [Fig. 1] \cite{12}. The present experiment with the piston-cylinder pressure cell confirmed the pressure-insensitive resistivity previously suggested by the experiment using the cubic anvil apparatus \cite{33}. Contrastingly, the resistivity of TTF-BA is as high as $\sim$1 $\mathrm{\Omega}$cm, which shows a little sensitivity to pressure in the entire pressure range studied. Remarkably, the resistivity values are 5-6 orders of magnitude different between the two systems at high pressures despite that both are commonly in the $\mathrm{I_{para}}$ phase.

%%%%%%%%%%%%%%%%%%%Fig2
\begin{figure*}
\centering
\includegraphics[width=17.5cm]{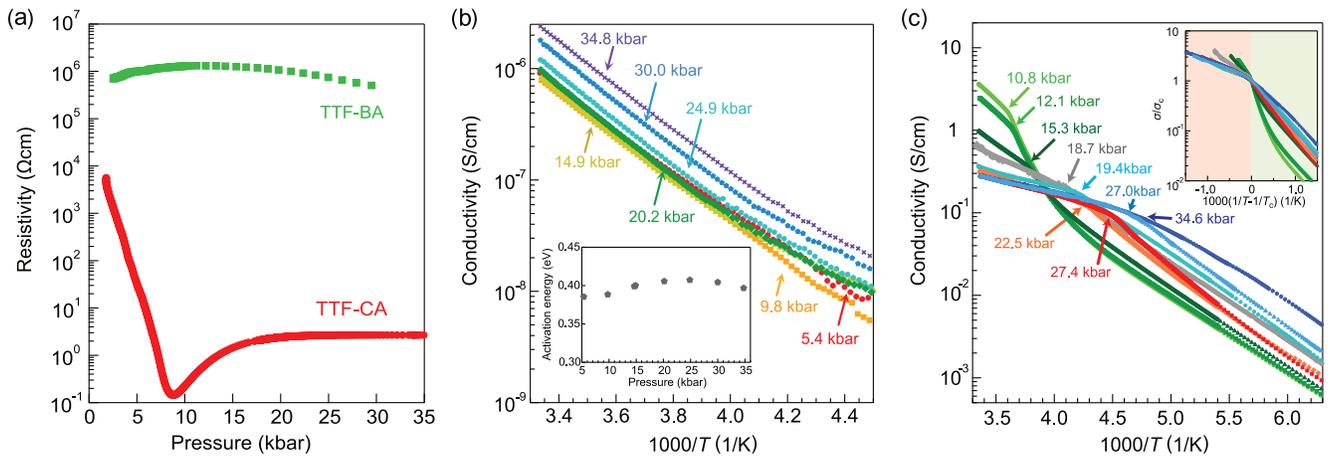}
\caption{(a) Comparison between the pressure dependence of the resistivities of TTF-CA along the $\textit{a}$-axis and TTF-BA along the $\textit{b}$-axis at room temperature. (b) Arrhenius plot of the resistivity of TTF-BA along the $\textit{b}$-axis at several pressures. The conductivity totally decreases with increasing pressure up to 10-15 kbar but increases with further pressure increase. The inset shows the pressure dependence of activation energy of TTF-BA estimated from the Arrhenius plot. (c) Arrhenius plot of the resistivity of TTF-CA along the $\textit{a}$-axis at several pressures. The arrows indicate the phase transition points, $\textit{T}_c$, between the $\mathrm{I_{para}}$ and $\mathrm{I_{ferro}}$ phases. The inset shows the Arrhenius plot of the normalized conductivity $\sigma$/$\sigma_c$ against 1000(1/$\textit{T}$-1/$\textit{T}_c$), where $\sigma_c$ is the  conductivity at $\textit{T}_c$. The red and green regions in the inset correspond to the $\mathrm{I_{para}}$ and $\mathrm{I_{ferro}}$ phases, respectively.
}
\label{Fig2} 
\end{figure*}
%%%%%%%%%%%%%%%%%%%Fig2

%\subsection{Temperature dependence of electrical resistivity of TTF-BA and TTF-CA}
We show the Arrhenius plots of the resistivities of TTF-BA and TTF-CA in Figs. 2 (b) and 2 (c), respectively. The resistivity of TTF-BA is characterized by the activation energies of 0.38-0.42 eV at every pressure up to 35 kbar [the inset of Fig. 2 (b)]. The charge soliton excitation expected in the NI transition materials is akin to a neutral single-molecule defect [Fig. 1], whose excitation energy is predicted to be an order of magnitude smaller than the charge-transfer (CT) energy gap \cite{13, 14, 15}. In TTF-BA, however, the charge gap, which is twice the activation energy, is nearly equal to the CT energy of $\sim$0.8 eV determined by the IR measurements \cite{34}, which is consistent with quasiparticle excitations over the Mott gap and rules out the solitonic excitations in TTF-BA. The small decrease in the charge gap above 25 kbar is ascribable to an increase of the transfer integral. On the other hand, the small increase up to 25 kbar may be due to the stabilization of the ionic state owing to a Madelung energy gain by lattice contraction.

The 5-6 orders of magnitude smaller resistivity in TTF-CA than in TTF-BA [Fig. 2 (a)] suggests that the $\mathrm{I_{para}}$ phase of TTF-CA should have much lower charge excitation energy than the CT energy or the excitation energy of band quasiparticles. As seen in Fig. 2 (c), the slope of the Arrhenius curve in the $\mathrm{I_{para}}$ phase gradually changes with pressure increased up to 20 kbar and is nearly unchanged above 20 kbar, which is more visible when the conductivity is normalized to the value at $\textit{T}_c$ as shown in the inset of Fig. 2 (c). Figure 3 (a) displays the activation energies estimated from the slope in the $\mathrm{I_{para}}$ phase, which is $\sim$0.26 eV at 10 kbar and drops to $\sim$0.10 eV at 20 kbar, then saturating to the range of 0.06-0.07 eV above 25 kbar.

This analysis, however, requires the following caution. The low resistivity of TTF-CA around 10 kbar [the red area in Fig. 3 (b)] is caused by the excitations of the NIDWs \cite{12}. As discussed in Ref. 12, the NI crossover region is inclined in the pressure-temperature ($\textit{P}$-$\textit{T}$) plane and the resistivity should be analyzed parallel to the tilted crossover line in the NIDW-active $\textit{P}$-$\textit{T}$ region. The activation energies determined in this way \cite{12}, which are reproduced in Fig. 3 (a), show large discrepancies with the above estimated activation energies below 20 kbar, thus, which should be taken as spurious values. On the other hand, the activation energy decreases to 0.06-0.08 eV in the pressure range above 20 kbar, being the same order with those around 10 kbar. These values are far smaller than a half of the CT excitation energy \cite{20} or the recently reported band gap, $\sim$0.35 eV \cite{38}, indicating the non-quasiparticle transport. The free NIDW excitations should be suppressed above 20 kbar far from the NI crossover pressure; thus, these results demonstrate the existence of another low-energy excitation carrier, that is, the soliton whose activation energy is theoretically predicted to be less than 0.1 eV \cite{13}. Overall, the activation energy of the NIDWs is considered to cross over to that of the solitons in a way depicted by the broad curve in Fig. 3 (a); the NIDW excitation energy taking a minimum value, 0.055 eV, at 9 kbar \cite{12} increases with pressure but gradually turns back to the similar value at high pressures though the soliton excitation energy should be over twice the NIDW excitation energy \cite{13, 15}. We speculate that local lattice deformation associated with the soliton formation may lower its creation energy; this is an issue of further investigation.

%\subsection{Analysis of the activation energy of charge soliton}
In what follows, we make quantitative discussion on the observed activation energy of 0.06-0.08 eV. The NI transition system has been modeled to the form of the Hubbard-type Hamiltonian that includes the on-site repulsive energy, $\textit{U}$, the inter-site attractive energy, -$\textit{V}$, and the site-alternating potential, $\mathit{\Delta}_0$, reflecting the energy difference between the HOMO of D molecule and the LUMO of A molecule \cite{39}. This Hamiltonian is reduced to the extended Hubbard model with $\textit{U}$, the repulsive $\textit{V}$, and the reduced $\mathit{\Delta}$ (=$\mathit{\Delta}_0$-4$\textit{V}$), and is further transformed to a phase Hamiltonian through the bosonization \cite{40}. The analytical solution of the energy of charge soliton, $\textit{E}_{\mathrm{CS}}$, was obtained by Fukuyama and Ogata \cite{14} as,
\begin{equation}
\textit{E}_{\mathrm{CS}}=\frac{2\textit{v}_{\rho}\sqrt{\gamma_c}}{\pi}(\cos \theta-\sin \theta (\frac{\pi}{2}-\theta))
\end{equation}
where $\textit{v}_{\rho}=2\textit{ta}+(\textit{U}+6\textit{V})\textit{a}/2\pi$, $\gamma_{\mathrm{c}}=(\textit{U}-2\textit{V})/\pi\textit{a}\textit{v}_{\rho}$, $\textit{a}$ is the lattice constant, and $\textit{t}$ is the transfer integral between the D and A molecules. $\theta$ is a phase variable for charge; for example, $\theta$ ($0\textless\theta\textless\pi/2$) is 0 and $\pi$/2 in the ionic and neutral limits, respectively.

The topological charge of the charge soliton is experimentally estimated by $\textit{q}_{\mathrm{CS}}/\textit{e}=\pm\rho_\mathrm{I}$, which is related to $\theta$ through $\textit{q}_{\mathrm{CS}}/\textit{e}=1-2\theta/\pi$ \cite{15}, where $\textit{e}$ is the elementary charge. The $\textit{q}_{\mathrm{CS}}$ value of 0.75$\textit{e}$ in the $\mathrm{I_ {para}}$ phase yields $\theta$=0.39 and thus $\cos \theta-\sin \theta (\frac{\pi}{2}-\theta)$ =0.47. The $\textit{v}_{\rho}$ and $\gamma_c$ are the functions of $\textit{t}$ and $\textit{V}$ that vary with pressure. Using the reported parameter sets of ($\textit{U}$, $\textit{V}$, $\textit{t}$) = (1.5, 0.7, 0.2) \cite{13} and (1.528, 0.604, 0.179) \cite{41} in eV at ambient pressure and taking account of the pressure effect \cite{11, 42}, we obtained $\textit{v}_{\rho}\gamma_{\mathrm{c}}^{1/2}$ at 20 kbar as 0.13 and 0.34, which yield $\textit{E}_{\mathrm{CS}}$ of 0.04 and 0.1 eV, respectively. The range of these values explains the experimental values of 0.06-0.08 eV, lending support to the view that the charge soliton dominates the electric conduction in the $\mathrm{I_ {para}}$ phase.

The solitonic electrical conduction is featured by little pressure dependence of activation energy as shown in Fig. 3 (a). $\rho_\mathrm{I}$ is known to be nearly pressure-independent in the $\mathrm{I_ {para}}$ phase \cite{23, 25, 27, 28, 31}, meaning that $\theta$ is so as well. The prefactor, $\textit{v}_{\rho}\gamma_{\mathrm{c}}^{1/2}$, in Eq. (1) is estimated to only vary by 1.6$\%$/kbar and 0.1$\%$/kbar for the two sets of ($\textit{U}$, $\textit{V}$, $\textit{t}$) values used in the above estimation \cite{13, 41} according to the pressure effect \cite{11, 42}. This explains the pressure-insensitive activation energy at high pressures. More rigorously, the electron-phonon coupling, the many-body effects of solitons, and the quantum fluctuations beyond the present consideration may influence the $\textit{E}_{\mathrm{CS}}$ value.

%%%%%%%%%%%%%%%%%%%Fig3
\begin{figure}
\includegraphics[width=9cm]{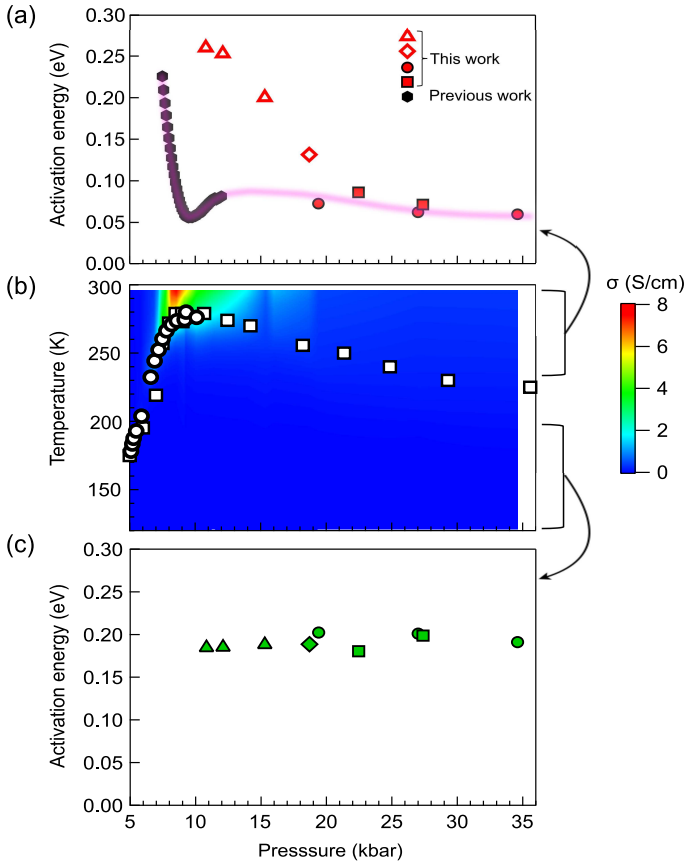}
\caption{(a) Pressure dependence of the activation energy of TTF-CA in the high temperature region above the feroelectric transition temperature, $\textit{T}_c$. The results of four samples measured are distinguished by the different symbols. The activation energies are obtained from the plots in Fig. 2 (c). The activation energies estimated along parallel to the inclined NI crossover line in the previous measurements \cite{12} are also displayed. The  pressure dependence of activation energy that apears to make sense is indicated by the guided curve. (b) Phase diagram of TTF-CA with the contour plot of electrical conductivity, $\sigma$, along the $\textit{a}$-axis extracted from Ref. 12. (c) Pressure dependence of activation energy of TTF-CA in the $\mathrm{I_{ferro}}$ phase.
}
\label{Fig3}
\end{figure}
%%%%%%%%%%%%%%%%%%%Fig3

%\subsection{Charge soliton density and diffusion coefficient}
Next, we discuss the diffusion dynamics of charge solitons. It is a consensus that the steady current of charge solitons requires the presence of spin solitons \cite{13, 14, 15}; thus, we take the effective charge of the carrier as $\pm\textit{e}$ that is the sum of the topological charges of the charge and spin solitons, $\textit{q}_{\mathrm{CS}}=\pm\rho_\mathrm{I}\textit{e}$ and $\textit{q}_{\mathrm{SS}}=\pm(1-\rho_\mathrm{I})\textit{e}$ \cite{15}. Theoretically, the excitation energy of the charge soliton is suggested to be higher than that of the spin soliton \cite{13, 14, 15}, and thus the number of thermally excited charge solitons should be less than that of spin solitons. Although both the charge and spin solitons are responsible for electrical conduction, it is governed by the number of minority carrier, that is, the charge soliton. Therefore, the electrical conductivity is expressed as $\sigma=\textit{e}\textit{n}_{\mathrm{CS}}\mu_{\mathrm{CS}}$, where $\textit{n}_{\mathrm{CS}}$  and $\mu_{\mathrm{CS}}$ are the density and mobility of charge soliton, respectively. Using the experimental value of $\sigma$=0.4 S/cm at 20 kbar, we obtain $\textit{n}_{\mathrm{CS}}\mu_{\mathrm{CS}}$ =2.5$\times10^{18}$ 1/Vscm. As the charge soliton is equivalent to a pair of combined NIDWs sandwiching a neutral molecule [Fig. 1], $\mu_{\mathrm{CS}}$ would be smaller than the mobility of the NIDW, which is estimated at $\mu_{\mathrm{NIDW}}$ =0.14 cm$^{2}$/Vs through $\sigma_\mathrm{{NIDW}}=(\textit{e}/2)\textit{n}_{\mathrm{NIDW}}\mu_{\mathrm{NIDW}}$ with $\sigma_\mathrm{{NIDW}}$=7 S/cm and $\textit{n}_{\mathrm{NIDW}}$$\sim$6.1$\times10^{20}$ 1/cm$^{3}$ (one soliton per $\sim$5 DA pairs) at 9 kbar \cite{12}. Assuming, $\textit{e.g.}$, $\mu_{\mathrm{CS}}=\mu_{\mathrm{NIDW}}/2\sim$0.072, we have the estimate of $\textit{n}_{\mathrm{CS}}\sim$3.5$\times10^{19}$ 1/cm$^{3}$ (one soliton per $\sim$88 DA pairs) at 20 kbar after a lattice contraction by 7$\%$ is considered \cite{42}. Then, the charge-soliton diffusion coefficient, $\textit{D}_{\mathrm{CS}}$, which is given by the Einstein’s relationship, $\mu_{\mathrm{CS}}=\textit{e}\textit{D}_{\mathrm{CS}}/\textit{k}_{\mathrm{B}}\textit{T}$, is $\sim$1.9$\times10^{-3}$ cm$^{2}$/s at 300 K, where $\textit{k}_{\mathrm{B}}$ is the Boltzmann constant. On the other hand, the diffusion coefficient of spin soliton, $\textit{D}_{\mathrm{SS}}$, was previously evaluated by $^{1}$H-NMR as 2.4$\times10^{-3}$ cm$^{2}$/s at 14 kbar at 300 K \cite{11}. It is surprising that the $\textit{D}_{\mathrm{CS}}$ and $\textit{D}_{\mathrm{SS}}$ values determined by completely different experimental methods nearly coincide with each other. This result indicates that the charge and spin solitons move together, strongly supporting the theoretical prediction that the spin solitons are required as well as the charge solitons to carry steady current \cite{13, 14, 15}.

%\subsection{Electrical conduction in the low-temperature phase}
As seen in Fig. 2 (c), the Arrhenius plot of resistivity in the $\mathrm{I_{ferro}}$ phase of TTF-CA appears approximately parallel for every pressure above 10 kbar and is characterized by the activation energies of 0.18-0.21 eV, which are about three times as large as the $\textit{E}_{\mathrm{CS}}$ values [Fig. 3 (c)]. Nevertheless, the activation energies are smaller than a half of the CT energy, 0.35 eV, suggesting that the charge carriers in the $\mathrm{I_{ferro}}$ phase are not the band quasiparticles either. In fact, the previous conductivity, NMR, and NQR studies of the $\mathrm{I_{ferro}}$ phase at 14 kbar suggested that the charge (and spin) carrier should be a polaron \cite{43}. The polaron is the combined excitation of a charge and a spin soliton, which does not break the three-dimensional ferroelectric order. The activation energy larger than that of the charge soliton is reasonable by considering the composite character of the polaron. According to the previous analysis, the mobility gap in the $\mathrm{I_{ferro}}$ phase, which also contributes to the activation energy in the polaron transport, was estimated at 0.02 eV \cite{43}. Thus, the observed activation energy of 0.18-0.21 eV in charge transport is nearly determined by the polaron creation energy. Its pressure-insensitivity reasonably accords with the pressure-insensitivity of the charge soliton activation energy.

%\section{Concluding remarks}
We have investigated the charge transport in the ionic phases of the donor-acceptor mixed-stack systems, TTF-CA and TTF-BA, under pressure. TTF-BA shows high resistivity of the order of $\sim$1 M$\Omega$cm (at room temperature) with activation energies of $\sim$0.4 eV, which are explained by the quasiparticle excitations over the Mott gap. In contrast, TTF-CA shows far lower resistivities of the order of $\sim$1 $\Omega$cm (at room temperature) and small activation energies of 0.06-0.08 eV in the $\mathrm{I_{para}}$ phase above 20 kbar, which are proved to be consistent with the charge soliton excitation energy evaluated with the theoretical model using the phase Hamiltonian. The excitation energy of NIDWs emerging at the lower pressures smoothly approaches the charge soliton excitation energy with increasing pressure, where the neutral domains progressively shrink to single neutral molecules, namely, the charge solitons. We have also estimated the diffusion coefficient of charge soliton and found it to roughly accord with that of spin soliton, suggesting that they move together. The present results provide foundations on the thermal excitations of topological defects in a quasi-1D ionic Mott-Peierls ferroelectric.

\acknowledgments
We thank M. Tsuchiizu, H. Seo, and H. Fukuyama for fruitful discussions. This work was supported by the JSPS Grant-in-Aids for Scientific Research (Grant Nos. JP18H05225, 19H01846, 20K20894, 20KK0060 and 21K18144), by CREST (Grant No. JPMJCR1661), Japan Science and Technology Agency. We also thank the Cryogenic Research Center at the University of Tokyo for supporting low-temperature experiments.

%\begin{references}

%\end{references}

\end{document}